\documentstyle[12pt]{article}
\topmargin - 1.0 cm

\begin{document}

\begin{titlepage}
\title{\bf OSCILLATOR AS A HIDDEN NON-ABELIAN MONOPOLE}
\author{
L.G. Mardoyan${}^1$,
A.N. Sissakian,
V.M. Ter-Antonyan${}^2$ \\[3mm]
{\small Joint Institute for Nuclear Research, }\\
{\small Dubna, Moscow Region, 141980, Russia}
}
\date{\nonumber}

\maketitle
\footnotetext[1]{E-mail:mardoyan@thsun1.jinr.dubna.su}
\footnotetext[2]{E-mail: terant@thsun1.jinr.dubna.su}

\begin{center}
\underline{Abstract}
\end{center}
{\small A non--Abelian $SU(2)$ model is constructed for a five--dimensional
bound system "charge--dyon" on the basis of the Hurwitz--transformed
eight--dimensional isotropic quantum oscillator. The principle of
dyon--oscillator duality is formulated; the energy spectrum and wave functions
of the system "charge--dyon" are calculated.}
\thispagestyle{empty}
\end{titlepage}
\newpage
\setcounter{page}1

Systems with nontrivial topology are the most interesting objects of
quantum mechanics [1,2], quantum field theory [3,4], and condensed matter
physics [5,6].

In pioneering papers [7,8], these systems were merely postulated; later on,
they were deduced in quantum field theory from the first principles [9,10].

The problem of generation of systems with nontrivial topology in quantum
mechanics was formulated in refs. [11, 12]; while in refs. [13,14], a
four--dimensional isotropic oscillator was related to an Abelian system
"charge--dyon".

This note is aimed at constructing a non--Abelian $SU(2)$ model that
describes the system "charge--dyon".

Our consideration is essentially based on the Hurwitz transformation
[15--18]:
\begin{eqnarray}
x_0 &=& u_0^2 + u_1^2 + u_2^2 + u_3^2
-u_4^2 - u_5^2- u_6^2 - u_7^2 \nonumber \\  [2mm]
x_1 &=& 2(u_0u_4 + u_1u_5 - u_2u_6 - u_3u_7) \nonumber \\  [2mm]
x_2 &=& 2(u_0u_5 - u_6u_4 + u_2u_7 - u_3u_6)  \\  [2mm]
x_3 &=& 2(u_0u_6 + u_1u_7 + u_2u_4 + u_3u_5) \nonumber \\  [2mm]
x_4 &=& 2(u_0u_7 - u_1u_6 - u_2u_5 + u_3u_4) \nonumber
\end{eqnarray}
Transformation (1) obeys the identity
$$
r\equiv (x_0^2 + ... + x_4^2)^{1/2} = u_0^2 + ... + u_7^2\equiv u^2
$$
We complement (1) with the angular coordinates [19]
\begin{eqnarray}
{\alpha}_1 &=& \frac{i}{2}\ln{\frac{(u_0 + iu_1)(u_2 - iu_3)}
{(u_0 - iu_1)(u_2 + iu_3)}} \nonumber \\ [2mm]
{\beta}_1 &=& 2\arctan{\left[\frac{(u_0 + iu_1)(u_0 - iu_1)}
{(u_2 + iu_3)(u_2 - iu_3)}\right]^{1/2}} \\ [2mm]
{\gamma}_1 &=& \frac{i}{2}\ln{\frac{(u_0 + iu_1)(u_2 + iu_3)}
{(u_0 - iu_1)(u_2 - iu_3)}} \nonumber
\end{eqnarray}
Transformations (1)--(2) transform the space
$R^8(\vec u)$ into the direct product of the space
$R^5(\vec x)$ and three--dimensional sphere  $S^3(\alpha_1,\beta_1,\gamma_1)$.
>From (2) it follows that
$\alpha_1 \in [0,2\pi),\,\,\, \beta_1 \in [0,\pi],\,\,\,
\gamma_1 \in [0,4\pi)$.

Consider, in $R^8(\vec u)$, the quantum oscillator
\begin{equation}
\frac{{\partial}^2\psi}{\partial u^2_\mu}
+ \frac{2M}{\hbar^2}\left(E-\frac{M\omega^2u^2}{2}\right)\psi=0,
\,\,\,\, u_{\mu}\in R^8
\end{equation}
Formulae (1)--(3) provide the basic result of our paper
\begin{equation}
\frac{1}{2M}\left(-i\hbar \frac{\partial}{\partial x_j}-
\hbar A^a_j \hat T_a\right)^2\psi + \frac{\hbar^2}{2Mr^2}
{\hat T}^2\psi - \frac{\alpha}{r}\psi = \epsilon \psi
\end{equation}
Here $j=1,2,...,5;a=1,2,3$; parameters $\epsilon$ and $\alpha$
are connected with the oscillator parameters
\begin{eqnarray*}
\epsilon = -M\omega^2 /8,\,\,\,\,\,\,\,\,\, \alpha = E/4
\end{eqnarray*}

Equations (3) and (4) are dual to each other.
In equation (3)
$\omega$ is fixed and $E$ is quantized, whereas in eq. (4)
$E$ is fixed and $\omega$ is quantized. These situations are alternative
to each other: that in eq. (3) having the meaning of energy acquires the
meaning of coupling constant in eq. (4), and vice versa.
For a fixed $E$ equation (3) selects an infinite set
of monoenergetic states with frequencies
$\omega_N=E/\hbar(N+4), N\in {\rm {\bf Z}}$. These states
form a complete system of functions describing the object
connected with eq.(4).

Further, ${\hat T}^2 = {\hat T_1}^2 + {\hat T_2}^2 + {\hat T_3}^2$,
and the operators $\hat T_a$ satisfy the commutation
relations
\begin{eqnarray*}
[\hat T_a,\hat T_b] = i\epsilon_{abc}\hat T_c
\end{eqnarray*}
and are explicitly written as follows:
\begin{eqnarray}
\hat T_1 &=& i\left(\cos{{\alpha}_1}\cot{{\beta}_1}
\frac{\partial}{\partial {\alpha}_1}+
\sin{{\alpha}_1}\frac{\partial}{\partial {\beta}_1} -
\frac{\cos{{\alpha}_1}}{\sin{{\beta}_1}}
\frac{\partial}{\partial {\gamma}_1}\right) \nonumber \\ [3mm]
\hat T_2 &=& i\left(\sin{{\alpha}_1}\cot{{\beta}_1}
\frac{\partial}{\partial {\alpha}_1} -
\cos{{\alpha}_1}\frac{\partial}{\partial {\beta}_1} -
\frac{\sin{{\alpha}_1}}{\sin{{\beta}_1}}
\frac{\partial}{\partial {\gamma}_1}\right)  \\ [3mm]
\hat T_3 &=& -i\frac{\partial}{\partial {\alpha}_1} \nonumber
\end{eqnarray}
The triplet of five--dimensional vectors $A_j^a$ is given by the expressions
\begin{eqnarray*}
A^1_j &=& \frac{1}{r(r + x_0)}(0, -x_4, -x_3, x_2, x_1)  \\ [2mm]
A^2_j &=& \frac{1}{r(r + x_0)}(0, x_3, -x_4, -x_1, x_2)  \\ [2mm]
A^3_j &=& \frac{1}{r(r + x_0)}(0, x_2, -x_1, x_4, -x_3)
\end{eqnarray*}
The vectors $A_j^a$ are orthogonal to each other,
\begin{eqnarray*}
A^a_jA^b_j = \frac{1}{r^2}\frac{(r - x_0)}{(r + x_0)}
{\delta}_{ab}
\end{eqnarray*}
and also to the vector $\vec x = (x_0,...,x_4)$.

Every term of the triplet coincides with the vector potential of
a 5--dimensional Dirac monopole with a unit magnetic charge and
the line of singularity extending along the negative axis
$x_0$, including point $x_0=0$.

Equation (4) describes a 5--dimensional system "charge--dyon".
The operators $\hat T_a$ represent components of the total momentum
operator of that system and are generators of the $SU(2)$ group of
invariance of equation (4). The term
${\hbar}^2{\hat T}^2/2Mr^2$ represents the correction
introduced by Goldhaber [20].

In the space $R^5(\vec x)$ we introduce the hyperspherical coordinates
\begin{eqnarray*}
x_0 &=& r\cos \theta  \nonumber \\ [2mm]
x_1 + ix_2 &=& r \sin \theta \cos \frac{\beta}{2}e^{i\frac{\alpha
+\gamma}{2}}  \\ [2mm]
x_3 + ix_4 &=& r \sin \theta \sin \frac{\beta}{2}e^{i\frac{\alpha -
\gamma}{2}}  \nonumber
\end{eqnarray*}
It is to be noted that
$\theta \in [0,\pi],\,\,\alpha \in [0,2\pi),\,\,\, \beta \in [0,\pi],\,\,\,
\gamma \in [0,4\pi)$.

Equation (4) in these coordinates assumes the form
\begin{equation}
\left(\Delta_{r \theta}
- \frac{{\hat J}^2}{r^2 \sin^2 \theta/2}
- \frac{{\hat L}^2}{r^2 \cos^2 \theta/2}\right)\psi +
\frac{2M}{\hbar^2}\left(\epsilon + \frac{\alpha}{r}\right)\psi=0
\end{equation}
where
\begin{eqnarray*}
{\hat J}^2 = {\hat J_1}^2 + {\hat J_2}^2 + {\hat J_3}^2,\,\,\,\,\,\,
{\hat L}^2 = {\hat L_1}^2 + {\hat L_2}^2 + {\hat L_3}^2,\,\,\,\,\,\,
\hat L_a = \hat J_a + \hat T_a
\end{eqnarray*}
and the operators $\hat J_a$ follow from formulae (5) under the change
$\alpha_1\rightarrow\alpha,\\
\beta_1 \rightarrow \beta, \gamma_1 \rightarrow \gamma$.
Then,
\begin{eqnarray*}
\Delta_{r \theta} =
\frac{1}{r^4}\frac{\partial}{\partial r}
\left(r^4 \frac{\partial}{\partial r}\right) +
\frac{1}{r^2 \sin^3 \theta}\frac{\partial}{\partial \theta}
\left(\sin^3 \theta \frac{\partial}{\partial \theta}\right)
\end{eqnarray*}

By $G$ we shall denote common eigenfunctions of the commuting operators
${\hat L}^2$ and ${\hat J}^2$ corresponding to the eigenvalues
$L(L+1)$ É $J(J+1)$.

If we look for the solution to eq. (4) in the form
\begin{eqnarray*}
\Psi=R(r)Z(\theta)G(\alpha,\beta,\gamma;\alpha_1,\beta_1,\gamma_1)
\end{eqnarray*}
then upon separating the variables $r$ and $\theta$ and introducing the
constant of separation $\lambda(\lambda+3)$  (for the time being
$\lambda$ is arbitrary), we arrive at two ordinary differential equations
whose solutions are given by the following expressions
\begin{eqnarray*}
Z(\theta)=(1-\cos\theta)^J(1+\cos\theta)^L \nonumber \\ [2mm]
{_2F}_1\left(-\lambda+J+L,\lambda+J+L+3,2J+2;\frac{1-\cos\theta}
{2}\right)
\end{eqnarray*}
\begin{eqnarray*}
R_{n_r\lambda}(r)=const
(2\kappa r)^\lambda e^{-\kappa r}
F\left(\lambda+2-\frac{1}{r_0\kappa}, 2\lambda+4; 2\kappa r\right)
\end{eqnarray*}
Here we introduced the notation:
\begin{eqnarray*}
r_0={\hbar}^2/M{\alpha}^2,\,\,\,\,\,\,\,\,\,\,\,\,
{\kappa}^2=-2M\epsilon/{\hbar}^2
\end{eqnarray*}
>From standard conditions it follows that
$\lambda$ should be either integer (including $\lambda = 0$)
or half--integer and the energy spectrum of the 5--dimen\-sional
bound system "charge--dyon" is of the form
\begin{equation}
\epsilon=-\frac{M\alpha^2}{2\hbar^2(n_r+\lambda+2)^2},\,\,\,\,\,\,
n_r \in {\rm {\bf Z}},\,\,\,\,\,\,\lambda \in {\rm {\bf Z}}
\end{equation}
The energy levels (7) are degenerate with the multiplicity
\begin{eqnarray*}
g_N=\frac{(N+7)!}{7!N!},\,\,\,\,\,\,\,\,\,\,\,\, N=2(n_r+\lambda)
\end{eqnarray*}

 Instead of an oscillator, one could consider a system with
the potential energy
\begin{eqnarray*}
M{\omega}^2u^2/2 + V(u^2)
\end{eqnarray*}
where $V(u^2)$ is an arbitrary function. Then, an extra term
$V(r)/4r$ would arise in equation (4), though equations (3) and (4) would
remain dual to each other.

There are no a priori reasons to consider that equation (3) is in a way
related to the problem of magnetic charge. To obtain a non--Abelian monopole
>from an oscillator, we should apply to the nontrivial system of
transformations (1)--(2); and therefore, we called the $8D$--oscillator
the hidden non--Abelian monopole.

\vspace{0.4cm}

We are grateful to L.Davtyan, A.Nersessian and G.Pogosyan
for discussions.

\end{document}